# Radio signature of a distant behind-the-limb CME on 2017 September 6

V.N.Melnik (1), H.O. Rucker (2), A.I. Brazhenko (3), M. Panchenko (4), A.A. Konovalenko (1), A.V. Frantsuzenko (3), V.V. Dorovskyy (1), M.V. Shevchuk (1)

(1) Institute of Radio Astronomy, National Academy of Sciences of Ukraine, Kharkov, Ukraine
(2) Commission for Astronomy, Austrian Academy of Sciences, Graz, Austria
(3) Gravimetrical Observatory, Poltava, Ukraine
(4) Space Research Institute, Austrian Academy of Sciences, Graz, Austria

**Abstract**

We discuss properties of a Type IV burst, which was observed on 2017 September 6, as a result of the powerful flare X 9.3. At decameter wavelengths this burst was observed by the radio telescopes STEREO A, URAN-2, and NDA at frequencies 5 - 35 MHz. This moving Type IV burst was associated with a coronal mass ejection (CME) propagating in the southwest direction with a speed of 1570 km/s. The maximum radio flux of this burst was about 300 s.f.u. and the polarization was more than 40%. In the frequency range of 8-33 MHz it continued for more than 2 hr. For STEREO A the associated CME was behind the limb, its longitudinal angle was about $160°$. This moving Type IV burst was observed by STEREO A at frequencies of 5-15 MHz in spite of the low sensitivity of STEREO A. This means that the radio emission directivity of a Type IV burst is rather wide. Assuming the plasma mechanism of Type IV radio emission we derived the plasma density distribution in the CME core at distances of 5.6 Rs and 9.8 Rs and its mass to be about $10^{16} g$. It is planned that the minimum perihelion of the Parker Solar Probe (PSP) spacecraft will be at about 9 Rs. So we discuss in what conditions PSP will be in if it crosses a similar CME core.

**Introduction**

Recently a Type IV burst observed on 2013 November 7 was discussed by Gopalswamy et al. (2016) and Melnik et al. (2018). This burst was associated with a coronal mass ejection (CME) moving into the direction of STEREO B; this was a limb event for STEREO A and a behind-the-limb one for ground-based radio telescopes. This moving Type IV burst was registered by STEREO B and by the radio telescopes URAN-2 (Ukraine) and the Nancay Decameter Array (NDA, France). The analysis of Type IV properties leads to the conclusion that the source of this Type IV radio emission was the core of the corresponding CME and the directivity of this radio emission was wide. Assuming the plasma mechanism of the Type IV burst, we found the plasma density distribution in the core, its mass, and magnetic field suitable confining the core plasma. The maximum radio flux of the Type IV burst at frequency 32 MHz was about 30-40 s.f.u., the polarization reached about 40 %, and the brightness temperature was about $10^9 K$. We have to investigate whether there are similar events, and what their properties are.

The powerful flare X 9.3 occurred in the active region NOAA 2673 at approximately 12:00 UT on 2017 September 6 (Figure 1). The CME associated with this flare was ejected in the southwest direction at a speed of 1570 km/s. Simultaneously, the radio telescope URAN-2 observed a powerful Type IV burst preceded by a group of Type

III and Type II bursts. The weak Type IV burst at frequencies lower than 40 MHz was registered by NDA, too. But it is important to note that STEREO A, for which this event was behind the limb, observed this Type IV burst up to 5 MHz. This means that even a radio telescope with low sensitivity can register a Type IV radio emission from a behind-the-limb CME, and in addition the directivity of this radio emission turns out to be rather wide, like for the 2017 November 7 event.

Data of the radio telescopes STEREO A, URAN-2, NDA, and the optical data of the Solar and Heliospheric Observatory (SOHO) and STEREO A have been used in this paper to analyze the properties of Type IV radio emission and the CME core, which is the source of its radio emission. We also discuss conditions of a possible passage of a similar CME core across the Parker Solar Probe (PSP) spacecraft at the time of its closest approach to the Sun.

**Observations**

The radio telescope URAN-2 (Poltava, Ukraine) has an effective area of 28,000 sq.m, and its operating frequencies are 8-33 MHz (Brazhenko et al., 2005, Konovalenko et al., 2016). The frequency-time resolution of observations was 4 kHz-100 ms (Ryabov et al., 2010) and these observations on 2017 September 6 were carried out from 4:56 to 14:20 UT.

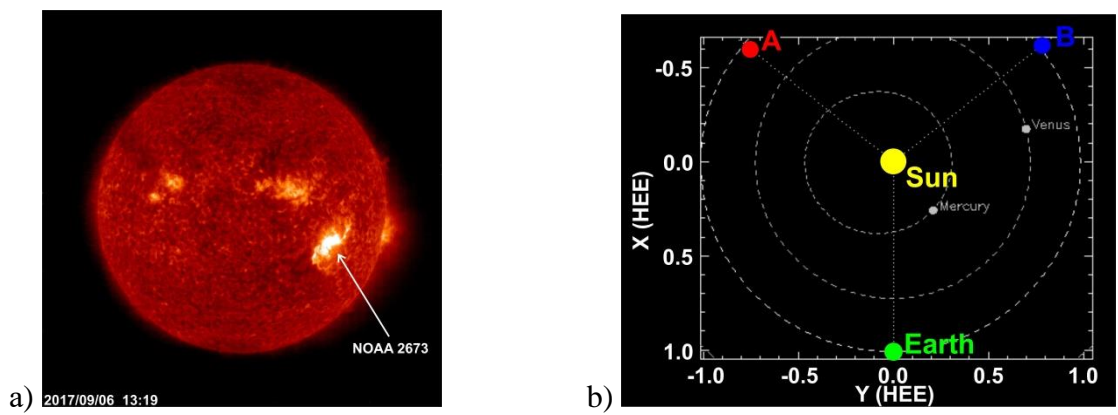

**Figure 1.** a) The solar disk according to SOHO and b) the location of STEREO A on 2017 September 6.

The dynamic spectrum of solar radio emission on this day is presented in the Figures 2 and 3. Prior to the Type IV burst a storm of Type III and Type IIIb bursts, with spikes, drift pairs, and S-bursts (from 4:56 to 12:00 UT) were observed. A powerful group of Type III bursts with fluxes of more than 1000 s.f.u. and low polarization, not higher than 20% (Figure 4), was observed immediately before the Type IV burst. X-ray observations (Lysenko et al., 2019) show that a maximum number of photons with energies of 20-80 keV was registered at 11:57-11:58 UT and namely at this time the first group of powerful Type III bursts was observed (Figure 4).

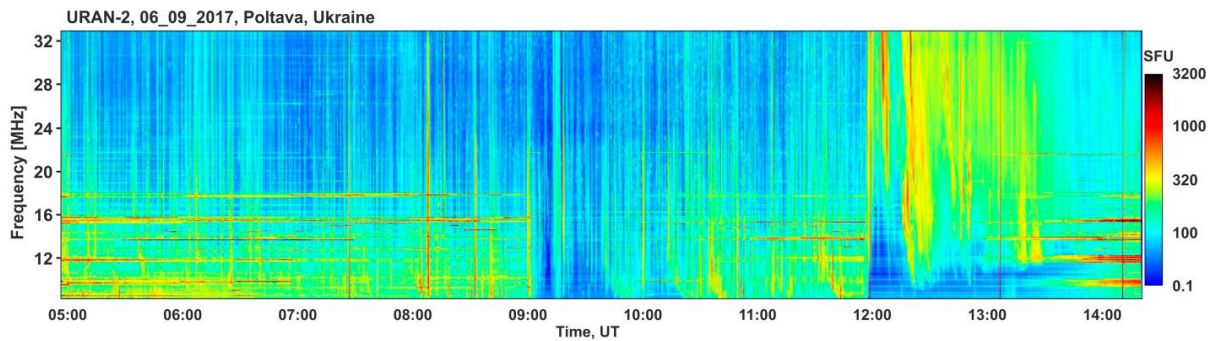

**Figure 2.** The dynamic spectrum of solar radio emission on 2017 September 6 according to URAN-2.

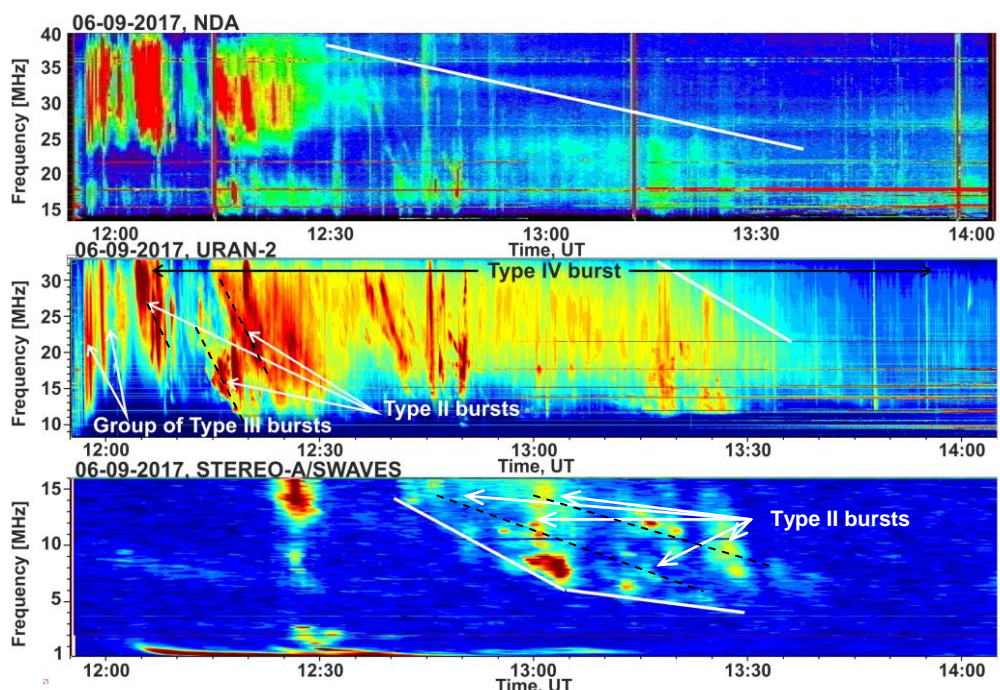

**Figure 3.** The dynamic spectra of Type IV burst, Type III bursts groups and Type II burst according to NDA, URAN-2 and STEREO A. Drifts of Type IV start and its end are marked with white lines. **Tracks of Type II bursts are marked with black dash-lines**.

Two harmonics of Type II bursts with fluxes of 300-1000 s.f.u. were observed directly after the powerful group of Type III bursts (Figures 3 and 4) at frequencies of 10-30 MHz. The drift rates of these Type II bursts were determined as -30 and -40 kHz/s for the fundamental and the harmonic, respectively, and a source of these bursts can be blast-shock initiated by flare. Their polarizations were 40% and 20%. Also, Type II bursts can be seen on the Type IV burst at frequencies of 6-15 MHz (Figure 3c) with drift rates of -2.9 and -3.3 kHz/s. These Type II bursts were probably initiated by a piston shock by the leading edge of the CME. The Type IV burst at a frequency of 32 MHz began at about 12:00 UT and continued to the end of that day's observations at about 14:20 UT. This burst yielded a maximum flux of about 300 s.f.u. at 12:10-12:20 UT. After that, the radio flux of the Type IV burst slowly decreased and had about 10 s.f.u. at the end of observations (Figure 4). The polarization maximum occurred after the flux maximum approximately at 13:30 UT reaching about 45%. Such peculiarities are standard for Type IV bursts (Melnik et al., 2008; 2018).

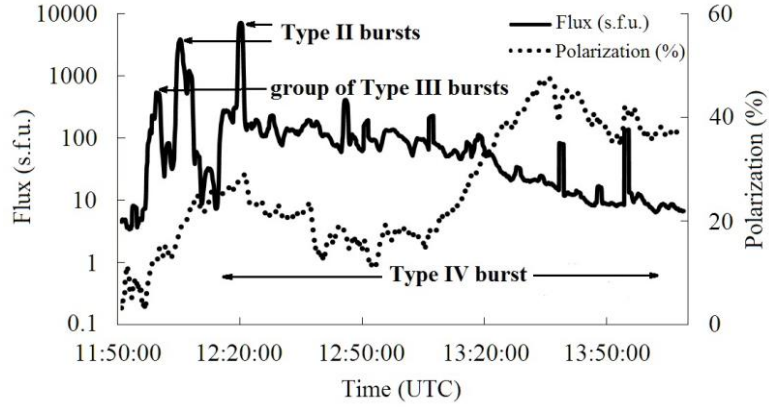

**Figure 4.** Profiles of flux and polarization of solar radio emission at frequency 32 MHz.

The drift of the start of Type IV at frequencies of 10-33 MHz was disguised by Type III bursts and by the fundamentals and the harmonics of Type II bursts (Figure 3b). The drift rates of the start of Type IV were about **-4** kHz/s at frequencies of 5-7 MHz and about -1.3 kHz/s at frequencies of 7-13 MHz (Figure 3c). The end of the Type IV burst drifted with a rate of about **-10** kHz/s at frequencies of 20-33 MHz. The Type IV burst had a fine structure in the form of Type III-like bursts with positive and negative drift rates (Figure 3). This is the usual situation for decameter Type IV bursts (Melnik et al., 2008; 2018). These bursts have drift rates and durations close to those for usual Type III bursts (Chernov et al., 2007; Antonov et al., 2014; Bouratzis et al., 2019; Dididze et al., 2019), but these bursts for most of the part were limited by the borders of the Type IV burst. Some bursts similar to Type II bursts were observed at frequencies of 15-33 MHz with drift rates of -50-80 kHz/s. These bursts can be radio emissions of different parts of the spherical shock initiated by the CME (Dorovskyy et al., 2018; Melnik et al., 2018; Zucca et al., 2018; Morosan et al., 2019).

The discussed Type IV burst was associated with the CME, initiated by the flare X 9.3 in the active region NOAA 2673 (Figure 1). This active region had a longitude of $\beta_1 = 30^0$ and a latitude of $10^\circ$. The CME moved in the southwest direction with a speed of 1570 km/s in the picture plane, and its real speed was 3140 km/s taking into account the longitudinal angle. Such a speed is one of the highest CME speeds (Gopalswamy et al., 2011). Both C2 (at 12:48 UT) and C3 (at 13:30 UT) observed by SOHO show CME cores (Figure 5). Assuming that the core is a sphere (Figure 6) we presented projections of the cores for those times. The core radius equals $1\,R_s$ and $1.7R_s$ at 12:48 UT and 13:30 UT, correspondingly and increasing in time. The centers of the core were at distances 2.8 and $4.9R_s$, respectively, in the picture plane.

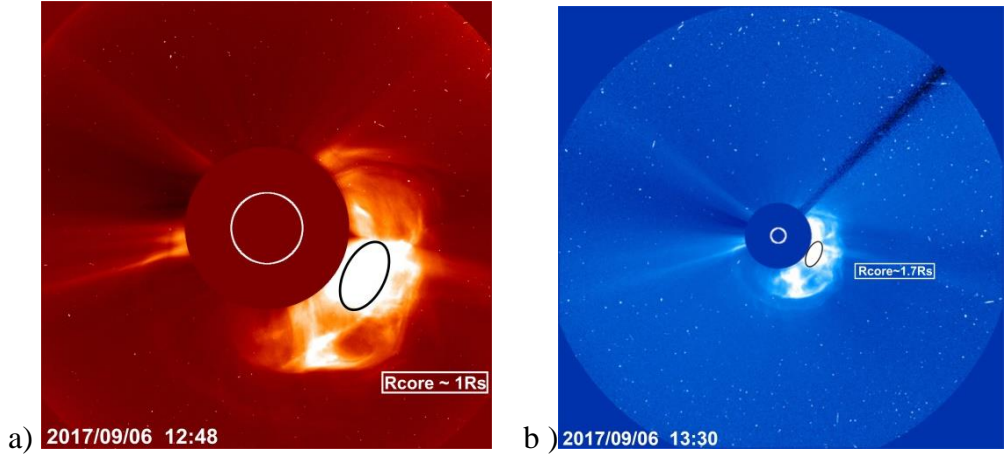

**Figure 5**. CME and their cores according to C2 and C3 at 12:48 and 13:30 UT.

**Discussion**

In Figure 6 the positions of the CME core at times $t_1 = 12:48\,\mathrm{UT}$ and $t_2 = 13:30\,\mathrm{UT}$ are presented, taking into account the longitudinal angle $\beta_1 = 30^0$ between the direction of the CME core motion and the direction to the Earth. The angle between the directions to the Earth and to STEREO A is $\beta_2 = 130^0$. Here the distances are $L_1 = 5.6 R_s$ and $L_2 = 9.8 R_s$. Considering that the CME appeared at $t_0 = 12:00\,\mathrm{UT}$ we find that the speed of the CME core was $V_1 = (L_1 - R_s)/(t_1 - t_0) \approx 1120\,km/s$ moving from the Sun to the distance $L_1$. Moving from $L_1$ to $L_2$ the speed was $V_2 = (L_2 - L_1)/(t_2 - t_1) \approx 1170\,km/s$. So the velocity of CME core practically did not change.

As already mentioned, the size of the core is increasing with time, from $R_1 = R_s$ to $R_2 = 1.7 R_s$ at distances $L_1$ and $L_2$, correspondingly. Following the paper by Melnik et al. (2018) we suppose that the plasma density in the core is exponentially distributed

$$n(r) = n_c \exp(-\alpha r),\tag{1}$$

where $n_c$ is the density in the center of the CME core. If the radio emission of the Type IV burst is generated by the plasma mechanism, then by escaping from the core's center at a maximum frequency of 33 MHz at the time $t_1$, the plasma density was $n_{33} = 1.34 \cdot 10^7\,cm^{-3}$. The radio emission at the minimum frequency of 15 MHz was generated at peripheral regions, where the density was $n_{15} = 2.77 \cdot 10^6\,cm^{-3}$. For the time $t_2$, when the maximum and minimum frequencies were 25 and 5 MHz, the corresponding densities were $n_{25} = 7.7 \cdot 10^6\,cm^{-3}$ and $n_5 = 3 \cdot 10^5\,cm^{-3}$. From the conditions that $n_{15} = n(R_1)$ and $n_5 = n(R_2)$, we derive the values $\alpha_1 = 1.6/R_s$ and $\alpha_2 = 1.9/R_s$ for the times $t_1$ and $t_2$, respectively. For these points of times the masses of the CME core

$$M = 4\pi m_p \int\limits_0^{R_{1,2}} n(r) r^2 dr\tag{2}$$

equal $M_1 = 1.10 \cdot 10^{16} g$ and $M_2 = 1.07 \cdot 10^{16} g$, respectively. This means that the mass of the core does not significantly change during the movement.

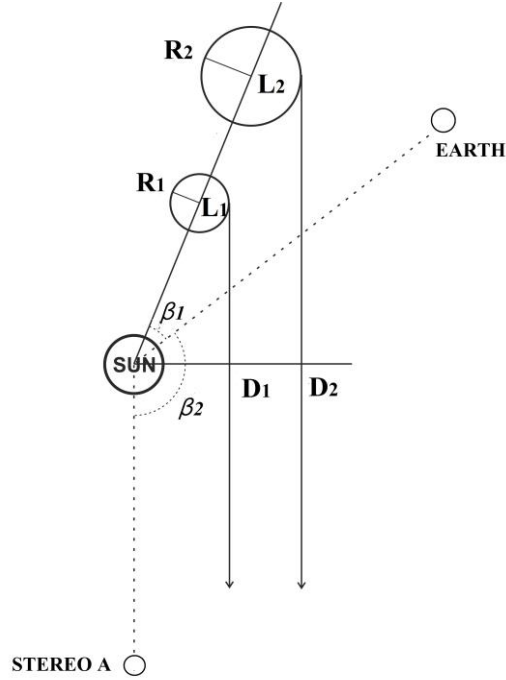

**Figure 6.** Locations of STEREO A, the Earth and the CME core at times $t_1$ and $t_2$ on 2017 September 6. Propagation of Type IV radio emission at 15 and 5 MHz towards STEREO A through the solar corona is shown.

The radio emission at frequencies of 15 and 5 MHz can be registered by STEREO A if the local plasma frequencies at $D_1 = L_1 \cos(\beta_1 + \beta_2 - 90^0) + R_1 = 2.92 R_S$ and $D_2 = L_2 \cos(\beta_1 + \beta_2 - 90^0) + R_2 = 5.1 R_S$ (Figure 6) are smaller than these registered frequencies. In Table 1 the local plasma frequencies in the corona models of Leblanc et al. (1998), Newkirk (Newkirk, 1961), Saito et al. (1970), Mann et al. (1999) and Vrsnak et al. (2004) are presented. We see that the radio emission at frequencies of 15 and 5 MHz can propagate through the solar corona in the direction to STEREO A and can be registered by the spacecraft. Talebpour & Pohjolainen (2018) and Pohjolainen & Talebpour (2020) assumed that this radio emission can be blocked by the shock between the core and STEREO A, but we see that some Type II bursts at these frequencies at corresponding times were absent. So we conclude that shocks did not block radio emission from the CME core.

Table 1. Local plasma frequencies at distances $D_1$ and $D_2$ in different corona models

| Solar Corona Model | $f_{pe}(D_1)$, MHz | $f_{pe}(D_2)$, MHz |
|---|---|---|
| Leblanc et al. | 4.3 | 1.4 |
| Newkirk | 10 | 4.9 |
| Saito et al. | 6 | 2.1 |

| Mann et al. | 6.83 | 2.48 |
|---|---|---|
| Vrsnak et al. | 12.4 | 3.8 |

According to the model by Mann et al. (1999), the coronal plasma densities at distances $L_1 = 5.6 R_s$ and $L_2 = 9.8 R_s$ are $n_{5.6 R_s} = 5.9 \cdot 10^4 cm^{-3}$ and $n_{9.8 R_s} = 2 \cdot 10^4 cm^{-3}$, correspondingly. Comparing these values with densities in peripheral regions of the CME core of $n_{15} = 2.77 \cdot 10^6 cm^{-3}$ and $n_5 = 3 \cdot 10^5 cm^{-3}$, we see that the ratios are $n_{15} / n_{5.6 R_s} \approx 47$ and $n_5 / n_{9.8 R_s} \approx 15$, in other words densities in the CME core are essentially higher than the density of the surrounding coronal plasma. The plasma in the CME core should be confined by the magnetic field. If the temperature in the core is $T = 10^5 K$, it needs $B = 3 \cdot 10^{-2} G$ and $B = 10^{-2} G$ when the CME core is located at $L_1 = 5.6 R_s$ and $L_2 = 9.8 R_s$, respectively. If the temperature in the core is assumed to be as high as $T = 10^6 K$, then these values should be a factor of 3 times higher. Similar magnetic field strengths for the confinement of the CME core plasma were needed in the case of the CME observed on 2013 November 7 (Melnik et al., 2018).

In the model of the decameter Type IV burst from the CME core, the frequency drift rate of the Type IV start is governed by the expansion of the CME core (Melnik et al., 2018). For our case the rate of core expansion is given by

$$V_{exp} \approx (R_2 - R_1) / (t_2 - t_1) = 200 km / s \tag{3}$$

This rate is smaller than that for the core expansion of the CME observed on 2013 November 7, when the expansion speed was 300 km/s. The drift rate of the Type IV start can be found by the equation

$$\frac{df}{dt} = \frac{f}{2n} \frac{dn}{dr} \frac{dr}{dt} = \frac{f}{2} (-\alpha) \frac{dr}{dt} \tag{4}$$

The calculated drift rate at 15 MHz is about -3.4 kHz/s and is -1.35 kHz/s at 5 MHz. These values are close to the observed rates of -4 and -1.3 kHz/s at these frequencies.

At the time 12:29 UT the radio fluxes are about 45 s.f.u. at a frequency of 15 MHz and its brightness temperature was about $T = 4 \cdot 10^9 K$. At the time 13:30 UT the radio flux was about 10 s.f.u. at a frequency of 12 MHz and the brightness temperature was $T = 2 \cdot 10^{10} K$. Such brightness temperatures can be understood when considering the plasma mechanism of radio emission of Type IV bursts.

In 2024 – 2025 the PSP spacecraft will approach the Sun at a distance of about $9 R_s$. In these years the solar activity will be high and practically every day four to five CMEs will be ejected into the solar corona, some of which may propagate and pass over the spacecraft. In these cases the spacecraft will be under the conditions of increased plasma density. If the spacecraft gets into the center of the CME core as discussed in this paper, then the gas-kinetic pressure of plasma $p = nkT$ in the center of the CME core will be $p_{CME} / p_{c.p.} = (n_c / n_{c.p.})(T_{CME} / T_{c.p.}) \approx 400$ ( $n_c = 7.7 \cdot 10^6 cm^{-3}$, $n_{9.8 R_s} = 2 \cdot 10^4 cm^{-3}$, $T_{CME} = T_{c.p.} = 10^6 K$ ) times larger then the pressure of the coronal plasma (c.p.) at these heights. Besides the gas-kinetic pressure of the CME, the pressure caused by the CME's speed $p = n_c M V_{CME}^2 / 2$ ( $n_c = 7.7 \cdot 10^6 cm^{-3}$ is the density in the center of the CME core, $M$ is the proton mass, $V \approx 1000 km / s$ is the speed of the CME core) will be even higher:

$p_{CME} / p_{c.p.} = (n_c / n_{c.p.})(MV_{CME}^2 / 2T_{c.p.}) \approx 2.5 \cdot 10^4$. So there are some risks for the PSP spacecraft in the case of "collision" with such CME cores.

**Conclusion**

The powerful flare X 9.3 on 2017 September 6 was associated with a CME, which was behind the limb for STEREO A (the longitude angle was $160^0$). This CME initiated a moving Type IV burst observed by URAN-2, NDA, and STEREO A in the frequency band of 5-35 MHz. The radio flux of the Type IV burst achieved 300 s.f.u. and its polarization was ~ 45%. Supposing that the CME core was the source of this Type IV burst, we found a density distribution in the core and its mass determined as about $1.1 \cdot 10^{16} g$ at distances of 5.6 and $9.8 R_S$ from the Sun. The density in the CME core was more than 100 times higher than the coronal density at the corresponding heights. The magnetic field strength necessary to confine the core plasma against expansion turned out to be about $10^{-1} \div 10^{-2} G$ at $5.6 R_S$ and $10^{-2} \div 10^{-3} G$ at $9.8 R_S$. It was shown that the drift rates of start of the CME is defined by the expansion of the CME core with a speed of about 200 km/s. The brightness temperature of the Type IV burst was $10^9 \div 10^{10} K$. This value can be understood in the frame of the plasma mechanism of radio emission.

It was also shown that if the PSP would be immersed within the CME core at perihelion, then the pressure of the core plasma would be more than 100 times higher than that of the coronal plasma at corresponding heights. The pressure of the CME wind with the speed of 1000 km/s will be higher by a factor of 25,000.

We thank the anonymous referee for useful comments that improved the manuscript. The research was supported by project "Spectr-3" of the National Academy of Sciences of Ukraine.

**ORCID iDs**

V. N. Melnik https://orcid.org/0000-0001-9239-6548
M. V. Shevchuk https://orcid.org/0000-0002-8275-9963